\documentstyle[preprint,aps,floats]{revtex}
\begin{document}
\draft
%@twocolumnfalse\endcsname
\title{A Herding Model with Preferential Attachment and Fragmentation}
\author{G.J. Rodgers$^{1,}$\thanks{e-mail: g.j.rodgers@brunel.ac.uk}
 and Dafang Zheng$^{1,2}$}
\address{$^{1}$ Department of Mathematical Sciences, Brunel  University,\\
 Uxbridge, Middlesex UB8 3PH, UK}
\address{$^{2}$ Department of Applied Physics, South China University of
Technology,\\ Guangzhou 510641, P.R. China} \maketitle

\maketitle
\thispagestyle{empty}

\begin{abstract}

We introduce and solve a model that mimics the herding effect in
financial markets when groups of agents share information. The
number of agents in the model is growing and at each time step
either (i) with probability $p$ an incoming agent joins an
existing group, or (ii) with probability $1-p$ a group is
fragmented into individual agents. The group size distribution is
found to be power-law with an exponent that depends continuously
on $p$. A number of variants of our basic model are discussed.
Comparisons are made between these models and other models of
herding and random growing networks.

\vspace*{0.2 true in} \noindent Keywords: Fragmentation;
Preferential Attachment; Herding; Financial Markets

\noindent PACS numbers: 02.50.cw, 05.40.-a, 89.75Hc.

\end{abstract}

% \narrowtext
%\vskip2pc]
%\begin{multicols}{2}

\newpage

%************************************************************
\section{Introduction}

Empirical studies of financial price-data on short time scales
have revealed that the price variations of various assets, indices
and currencies have fat-tails \cite{benoit,book}. These tails have
been shown to be power-law with an exponent in the distribution of
returns close to 4 in a number of different markets
\cite{stanley,dacor}. Beyond this power-law behaviour the
distribution crosses over to an exponential decay or to a steeper
power-law \cite{stanley}.

It is believed that this behaviour is brought about by a herding
effect in which groups of agents all behave in the same way. Cont
and Bouchaud \cite{CB} introduced a model of randomly connected
agents to investigate this herding effect. In \cite{CB} agents are
connected with probability $p$ and agents that are part of the
same group share information and make the same decisions. The
parameter $p$ is tuned to be close to the percolation threshold in
order to obtain a power-law distribution of group sizes. This in
turn leads to a power-law distribution of returns.

In \cite{EZ} an extension to this picture was introduced in which,
instead of being static, the network of agents evolves dynamically
as decisions are being made or as agents exchange information. In
this model, which was solved exactly in \cite{rodgers}, at each
time step  either two groups of agents shared their information
and were aggregated, or a group of agents trade, using their
information, and the group is then fragmented.

In this paper we introduce an alternative kinetic model for this
phenomenon. In our model, as in \cite{EZ,rodgers}, when a group of
agents trade, they use up their shared information, and are {\it
fragmented} to become individual agents again. However, we allow
new agents to enter the system and join the existing groups. When
this occurs the new agents join a group of size $k$ with a rate
proportional to $k$, so that large groups grow more quickly than
small ones. This seems reasonable as one would expect a new agent
to be more likely to come into contact with a member of a large
group than with a member of a smaller group. This process is
called {\it preferential attachment} in the theory of random
growing networks where an incoming node is more likely to connect
to a node with a high degree \cite{ba,sid}.

In Sec. 2 we present our model and find an exact analytical
expression for the group size distribution. In Sec. 3 we introduce
a generalised version of our initial model and a few particular
cases are solved. We discuss our work and draw some conclusions in
Sec. 4.

%************************************************************

\section{The Model}

\label{sec:model}

We introduce a model in which at each time step one of two events
can occur. With probability $p$ an agent is added to the system
and joins a group of size $k$ with a rate proportional to $k$.
Alternatively, with probability $q=1-p$ a group is selected at
random, with a rate independent of the size of the group, and the
group is fragmented into individual agents. Consequently the
number $n_{k}(t)$ of groups of size $k>1$ at time $t$ evolves like
\begin{eqnarray}
\label{rate} \frac{dn_{k}(t)}{dt}=
\frac{p}{M(t)}\left[(k-1)n_{k-1}-kn_{k}\right]
-q\frac{n_{k}}{N(t)}
\end{eqnarray}
and the number of groups with only one agent, or equivalently the
number of agents in a group on their own, behaves like
\begin{eqnarray}
\label{one} \frac{dn_{1}(t)}{dt}= -p\frac{n_{1}}{M(t)}
+\frac{q}{N(t)}\sum_{k=2}^{\infty}{kn_{k}}.
\end{eqnarray}
In these equations
\begin{equation}
\label{groups}
 N(t)=\sum_{k=1}^{\infty}n_{k}(t)
\end{equation}
represents the number of groups and
\begin{equation}
\label{agents}
 M(t)=\sum_{k=1}^{\infty}kn_{k}(t)
\end{equation}
is the number of agents in the system. The first term on the right
hand side of Eq.(\ref{rate}) describes the addition of a new agent
to an existing group and the last term describes the fragmentation
of a group of size $k$ into $k$ groups of size 1. In
Eq.(\ref{one}) the first term on the right hand side is the
destruction of free agents caused by the arrival of a new agent
and the second, summation, term is the creation of individual
agents by the fragmentation of other groups. Using rate equations
(\ref{rate}) and (\ref{one}) it is a simple matter to show that
\begin{equation}
\label{dn}
 \frac{dN(t)}{dt}=(1-p)\left[\frac{M(t)}{N(t)}-1\right]
\end{equation}
and
\begin{equation}
\label{dm}
 \frac{dM(t)}{dt}=p.
\end{equation}
Equation (\ref{dn}) represents the fact that with probability
$1-p$, {\it on average}, the number of groups increases by the
average group size minus one. Similarly, Eq.({\ref{dm}) indicates
that with probability $p$ the number of agents increases by $1$.
The form of Eqs.(\ref{rate},\ref{one},\ref{dn},\ref{dm}) suggests
that the solution for $n_{k}(t),$ for $k=1,2,...,$ is linear in
time for large t. In this limit we can solve
Eqs.(\ref{dn},\ref{dm}) to yield
\begin{equation}
N(t)=\alpha t \qquad \hbox{and} \qquad M(t)=pt
\end{equation}
where
\begin{equation}
\label{alpha}
 \alpha =
\frac{1-p}{2}\left[\sqrt{4\frac{p}{1-p}+1}-1\right].
\end{equation}
Writing
\begin{equation}
n_{k}(t)=tc_{k}
\end{equation}
we find that for $k>1$
\begin{equation}
\label{iter}
c_{k}=\left[(k-1)c_{k-1}-kc_{k}\right]-\frac{1-p}{\alpha}c_{k}.
\end{equation}
Using an initial condition obtained from Eq.(\ref{one}) we can
solve Eq.(\ref{iter}) to give
\begin{equation}
c_{k}=\frac{p(1-p)}{\alpha} \Gamma (\beta)\frac{\Gamma
(k)}{\Gamma(k+\beta)}
\end{equation}
where
\begin{equation}
\beta =2+\frac{1-p}{\alpha}=2
\left[\frac{\sqrt{4\frac{p}{1-p}+1}}{\sqrt{4\frac{p}{1-p}+1}-1}\right].
\end{equation}
As $k \to \infty$,
\begin{equation}
c_{k}\sim k^{-\beta}.
\end{equation}
Varying $p$ in $[0,1]$, we find that $\beta$ can take any value
$\beta >2$. As $p \to 0$ then $\beta \to \infty$ and when $p \to
1$ then $\beta \to 2$. At $p=q=1/2$,
$\beta=(\sqrt{5}+1)\sqrt{5}/2$.

 The distribution of returns,
$R(k)$, is the distribution of the relative difference between the
number of buyers and the number of sellers. This can be obtained
by realising that a group of agents of size $k$ trades with rate
$kc_{k}$. If we assume that the traded amount is proportional to
the number of agents in the group, then we find that
\begin{equation}
\label{return}
 R(k) \sim kc_{k} \sim k^{-\delta}
\end{equation}
where $\delta = \beta -1$. By varying $p$ we can allow $\delta$ to
take any value greater than $1$.

%************************************************************

\section{Generalisations}

The above models can be generalised by allowing incoming agents to
join groups of size $k$ with rate $A_{k}$ and fragmenting groups
of size $k$ with rate $B_{k}$. In the previous section we
considered the model with $A_{k}=k$ and $B_{k}=1$. The rate
equations (Eqs.(\ref{rate},\ref{one})) can be rewritten
\begin{eqnarray}
\label{rate1}
\frac{dn_{k}(t)}{dt}=
\frac{p}{A(t)}\left[A_{k-1}n_{k-1}-A_{k}n_{k}\right]
-qB_{k}\frac{n_{k}}{B(t)}
\end{eqnarray}
and
\begin{eqnarray}
\label{one1}
\frac{dn_{1}(t)}{dt}= -pA_{1}\frac{n_{1}}{A(t)}
+\frac{q}{B(t)}\sum_{k=2}^{\infty}{kB_{k}n_{k}}.
\end{eqnarray}
In these equations
\begin{equation}
\label{a}
 A(t)=\sum_{k=1}^{\infty}A_{k}n_{k}(t)
\end{equation}
and
\begin{equation}
\label{b}
 B(t)=\sum_{k=1}^{\infty}B_{k}n_{k}(t).
\end{equation}
The terms in Eqs.(\ref{rate1},\ref{one1}) have the same meaning as
those in Eqs.(\ref{rate},\ref{one}). Obviously we cannot solve the
above equations for general $A_{k}$ and $B_{k}$ so instead we
consider four simple special cases.

{\bf Model A}\quad $A_{k}=B_{k}=1$. This is the simplest model in
this class. With constant interaction kernels one would expect the
power-law behaviour to disappear. This is indeed what happens and
we easily find that
\begin{equation}
c_{k}=(1-p){\left[\frac{p}{\alpha +1}\right]}^{k}
\end{equation}
with $\alpha$ given in Eq.(\ref{alpha}).

{\bf Model B}\quad $A_{k}=k+\lambda$ and $B_{k}=1$. Here we keep
the fragmentation rate constant but adjust the preferential
attachment by adding a constant $\lambda > -1$ to the rate. As in
the work on random growing networks, \cite{ba}, we retain a
power-law group size distribution, but with a modified exponent
\begin{equation}
\beta = 2\frac{\sqrt{4\frac{p}{1-p}+1}+\lambda}
{\sqrt{4\frac{p}{1-p}+1}-1}.
\end{equation}
As before, varying $p$ in $[0,1]$ gives values of $\beta$ between
$2$ and $\infty$.

{\bf Model C}\quad $A_{k}=B_{k}=k$. In this model the preferential
attachment and the fragmentation proceed at the same rate and we
find that
\begin{equation}
c_{k} \sim \frac{\Gamma (k)}{\Gamma (k+p+1)} p^{k}
\end{equation}
so that the distribution is a power-law with an exponential
cut-off for large $k$.

{\bf Model D} \quad $A_{k}=1$ and $B_{k}=1/k$. Here the ratio
$A_{k}/B_{k} \sim k$ as in our first model. Hence one might
anticipate that the group size distribution would be power-law. In
fact
\begin{equation}
c_{k}=c_{1} \frac{\Gamma (k+1)\Gamma(2+\frac{\Delta +1-p}{\Delta
+p})}{\Gamma (k+\frac{\Delta +1-p}{\Delta +p}+1)}
{\left[\frac{p}{\Delta +p}\right]}^{k-1}
\end{equation}
where $\Delta (p)$ is determined by
\begin{equation}
\Delta = \sum_{k=1}^{\infty}{c_{k}}.
\end{equation}
As in Model C, this version exhibits a power-law with an
exponential cut-off. The cut-off remains for all values of $p$.

%************************************************************

\section{Discussion}

We have introduced and solved a kinetic model of herding which
exhibits a power-law distribution of group sizes for all its
parameter values. The exponent of the power-law can be varied
continuously. Other systems introduced to investigate the herding
phenomenon, \cite{CB,EZ,rodgers}, required a tuning of their
parameters to obtain a power-law. Furthermore, the exponent
obtained in \cite{CB,EZ,rodgers} is much smaller than those
obtained empirically for this phenomenon
\cite{book,stanley,dacor}. We would require a value of $p \approx
1/3$ to give a distribution of returns with an exponent $\approx
4$, Eq.(\ref{return}), to match the empirical results
\cite{book,stanley,dacor}.

Our system is open and growing with new agents are continually
entering the system. These new agents join an existing group with
preferential attachment. This is in contrast to \cite{EZ,rodgers},
where there were a fixed number of agents and the groups increased
in size by coagulation. In this sense our model is reminiscent of
random growing systems \cite{ba,sid}, where power-law
distributions are often found. We have been able to make use of
similar solution techniques to those employed on network models
\cite{sid}.

A consideration of a few variants of our model indicates that a
delicate balance between fragmentation and attachment is required
to obtain power-laws over a range of parameter values. An
exponential group size distribution is found for the constant
coefficient model and a power-law mediated by an exponential
cut-off is obtained when $A_{k}=A_{k}=k$ and $A_{k}=1, B_{k}=1/k$.
It is only when we keep the same rate of fragmentation, and
enhance the preferential attachment with an additive constant,
that power-laws are again recovered. This suggests that
preferential attachment is a necessary condition for power-laws in
these systems.

%*************************************************************

\section{Acknowledgements}

We would like to thank the Leverhulme Trust and the China
Scholarship Council for their financial support.
 %*********************FIGURES******************************

%************************************************************

%\end{multicols}

\begin{thebibliography}{99}

\bibitem{benoit} B. Mandelbrot, J. Business {\bf 36} 394 (1963).

\bibitem{book} R.N. Mantegna and H.E. Stanley, {\it An Introduction to
Econophysics: Correlations and Complexity in Finance,} (Cambridge
University Press, Cambridge, 2000).

\bibitem{stanley} P. Gopikrishnan, M. Meyer, L.A.N. Amaral
and H.E. Stanley, Eur. Phys. J. B {\bf 3} 139 (1998).

\bibitem{dacor} M.M. Dacorogna, U.A. Muller, R.J. Nagler,
R.B. Olsen and O.V. Pictet, J. Inter. Money and Finance {\bf 12}
413 (1993).

\bibitem{CB}R. Cont and J.-P. Bouchaud, Macroeconomic Dynamics {\bf 4} 170 (2000).

\bibitem{EZ} V.M. Egu\'{\i}luz and M.G. Zimmermann,
Phys. Rev. Lett. {\bf 85} 5659 (2000).

\bibitem{rodgers} R. D'Hulst and  G.J. Rodgers,
Int. J. Theor. Appl. Finance {\bf 3} 609 (2000).

\bibitem{ba} R. Albert and A.-L. Barab\'{a}si, preprint cond-mat/0106096.

\bibitem{sid} P.L. Krapivsky and S. Redner, Phys. Rev. E {\bf 63} 066123
(2001).
\end{thebibliography}
\end{document}